\documentclass[twocolumn,pra,showpacs,superscriptaddress]{revtex4-1}
\usepackage{amsfonts}
\usepackage{amssymb}
\usepackage{amsmath}
\usepackage{graphicx}
\usepackage{epsfig}
\usepackage{color}

\begin{document}

\title{Asymmetric lasing at spectral singularities}
\author{L. Jin}
\email{jinliang@nankai.edu.cn}
\affiliation{School of Physics, Nankai University, Tianjin 300071, China}

\begin{abstract}
Scattering coefficients can diverge at spectral singularities. In such
situation, the stationary solution becomes a laser solution with outgoing
waves only. We explore a parity-time ($\mathcal{PT}$)-symmetric
non-Hermitian two-arm Aharonov-Bohm interferometer consisting of three
coupled resonators enclosing synthetic magnetic flux. The synthetic magnetic
flux does not break the $\mathcal{PT}$ symmetry, which protects the symmetric transmission. The features and conditions of symmetric,
asymmetric, and unidirectional lasing at spectral singularities are
discussed. We elucidate that lasing affected by the interference is
asymmetric; asymmetric lasing is induced by the interplay between the synthetic magnetic flux and the system's non-Hermiticity. The
product of the left and right transmissions is equal to that of the
reflections. Our findings reveal that the synthetic magnetic
flux affects light propagation, and the results can be applied in the design
of lasing devices.
\end{abstract}

\maketitle

%\pacs{42.25.Bs, 42.60.Da, 11.30.Er}

%
% ma

%\title{Asymmetric lasing at the spectral singularity of a parity-time symmetric Aharonov-Bohm interferometer}

%42.25.Bs	Wave propagation, transmission and absorption [see also 41.20.Jb¡ªin electromagnetism; for propagation in atmosphere, see 42.68.Ay; see also 52.40.Db Electromagnetic (nonlaser) radiation interactions with plasma and 52.38-r Laser-plasma interactions¡ªin plasma physics]
%42.55.Ah	General laser theory
%42.60.Da	Resonators, cavities, amplifiers, arrays, and rings
%42.25.Hz	Interference
%03.65.Vf
%11.30.Er
%%%%%%%%%%%%%%%%%%%%%%%%%%  body  %%%%%%%%%%%%%%%%%%%%%%%%%%

\section{Introduction}

The theory of parity-time ($\mathcal{PT}$)-symmetric systems has been
extensively investigated~\cite%
{Bender98,Znojil,Ali02,Heiss,Ruschhaupt,DNC08,DNC2008,NM2008,LonghiPRL,TK2009,Kottos2010,HangPRL,HJing,LV,YuxiLiu}%
. In experiments, the use of optical platforms is fruitful for investigating
$\mathcal{PT}$-symmetric non-Hermitian quantum mechanics. To construct $%
\mathcal{PT}$-symmetric systems, passive and active devices are mostly used.
In passive systems, elements have varying loss rates~\cite{Observe}. A
passive system is equivalently described by a $\mathcal{PT}$-symmetric
Hamiltonian after an overall loss rate has been removed; however, mode
amplification is impossible in passive systems. In active systems, the gain
balances the loss to form a $\mathcal{PT}$-symmetric system~\cite%
{CERuter,PengNP,PengScience,Chang,Zhang}. Under steady state conditions, a
resonator doped with erbium ions under pumping induces a gain that balances
the loss in the dissipative resonator. Active $\mathcal{PT}$ symmetric
devices have numerous applications such as in optical isolators~\cite{Chang}
and single-mode $\mathcal{PT}$-symmetric lasing~\cite{FengScience,Hodaei}.

The non-Hermiticity induces asymmetric and nonunitary scattering in non-Hermitian systems~\cite{Kalish,AliM}. Numerous intriguing phenomena such as unidirectional reflectionless~\cite{Nature2012,NatMaterFeng}, invisible cloaking~\cite{ZLin,Alu}, asymmetric and robust light transport have been reported~\cite%
{SLonghi2015SR,Zhao2015SR}. Spectral singularities exist in the non-Hermitian scattering system, where the scattering coefficient diverges and the completeness of eigenstates is spoiled~\cite{AliPRL,SSlonghi,Ahmed2011,Main,Liu}. At spectral singularities, coherent perfect absorption~\cite{CPA,Science2011,CPAexp} and $\mathcal{PT}$-symmetric laser absorber~\cite{SL82,YDChong106,Hang} have been demonstrated.

In Hermitian systems, symmetric scattering exists even in the presence of synthetic magnetic flux, where photons mimic electrons in magnetic field~\cite{Flux1,Flux2,Flux3,Flux4,Topo1,Topo2,Topo3}. The synthetic magnetic flux induces asymmetric couplings but does not break the time-reversal symmetry. The time-reversal symmetry protects the optical reciprocity and the symmetric scattering.
However, in non-Hermitian systems, the synthetic magnetic flux and the non-Hermiticity can induce novel asymmetric behaviors~\cite{JLSR}.
The synthetic magnetic flux enclosed in a loss resonator side-coupled to a resonator chain
helps to create a unidirectional perfect absorber in which light can avoid
occupation of the loss resonator. The proposed system has full
reflectionless absorption on one direction, and full reflectionless
transmission in the opposite direction~\cite%
{SLonghi2015OL,LXQ,RamezaniUPA,JLSR2}. In contrast to a perfect absorber, a $%
\mathcal{PT}$-symmetric laser at non-Hermitian spectral singularities, where
symmetrical lasing toward both sides of a scattering system occurs has been
demonstrated~\cite{SL82,YDChong106,PWang}. Moreover, unidirectional lasing
at unidirectional spectral singularities has been proposed~\cite{USS}.

In this paper, we focus on a $\mathcal{PT}$-symmetric non-Hermitian system
consisting of three coupled optical resonators and in which synthetic
magnetic flux is enclosed. The three coupled resonators are embedded in a
resonator array and collectively serve as an Aharonov-Bohm (AB) interferometer. The
non-Hermiticity of the system is due to the balanced gain and loss in the
resonators. Synthetic magnetic flux is induced through an asymmetric coupling between neighboring resonators. $\mathcal{PT}$
symmetry holds in the presence of synthetic magnetic flux and
enables symmetric light transmission even though the coupling
is asymmetric. The scattering coefficients diverge at
spectral singularities of the non-Hermitian interferometer, with the
scattering wave function in the steady state representing lasing. The
features and conditions of symmetric, asymmetric, and unidirectional lasing
at spectral singularities are discussed in this paper. The synthetic magnetic flux affects the light interference and
leads to asymmetric lasing in the interferometer. As an illustration, we
evaluate the scattering properties of a uniformly coupled resonator system.
At spectral singularities, lasing asymmetry is minimal at trivial synthetic magnetic flux and maximal at half-integer synthetic magnetic flux. Wave emission has the signature of
spectral singularities in the wave packet scattering process.

The remainder of this paper is organized as follows. In Sec.~\ref{Model}, we
model the $\mathcal{PT}$-symmetric coupled resonator AB interferometer. In
Sec.~\ref{AsymmetricLasing}, we describe its spectral singularities and
asymmetric lasing. In Sec.~\ref{EG}, we consider a uniformly coupled
resonator as an illustration. Finally, we summarize our study in Sec.~\ref%
{Summary}.

\section{$\mathcal{PT}$-symmetric AB interferometer}

\label{Model}
Photons do not directly interact with magnetic fields; however, the photonic
analogy of the AB effect is realized through various methods
such as dynamic modulation of material permittivity, photon-phonon
interaction, and magneto-optical effects~\cite{Flux1,Flux2,Flux3,Flux4}. An AB interferometer is depicted in Fig.~\ref{fig1}. The
ring-shaped resonators are the primary resonators; they are coupled through
the stadium-shaped auxiliary resonators~\cite{HafeziIJMP}. The AB
interferometer consists of two uniformly coupled passive ring resonator
arrays and three ring resonators in the center. The primary resonator
frequency is $\omega _{\mathrm{c}}$. Resonator $+1$ has loss (green in Fig.~%
\ref{fig1}), resonator $-1$ has gain (pink in Fig.~\ref{fig1}), and
resonator $0$ is passive (white in Fig.~\ref{fig1}). The resonators are
evanescently coupled in a ring configuration and embedded in the resonator
array between resonators $+2 $ and $-2$. The coupling of the uniform
resonator array is $-\kappa $. The loss is caused by resonator dissipation,
whereas the gain is induced by pumping the ions doped in resonator $-1$. The
gain is modeled by a constant rate $\gamma $ when the gain is not close to
saturation in the linear region~\cite{PengNP,FengScience}. The couplings
between resonators $0$ and $\pm 1$ are $-g$. {%
The resonators $-1$, $0$, and $+1$ compose the scattering center $H_{\mathrm{c}}$.

\begin{figure}[tbp]
\centering\includegraphics[bb=0 0 445 140, width=8.8 cm, clip]{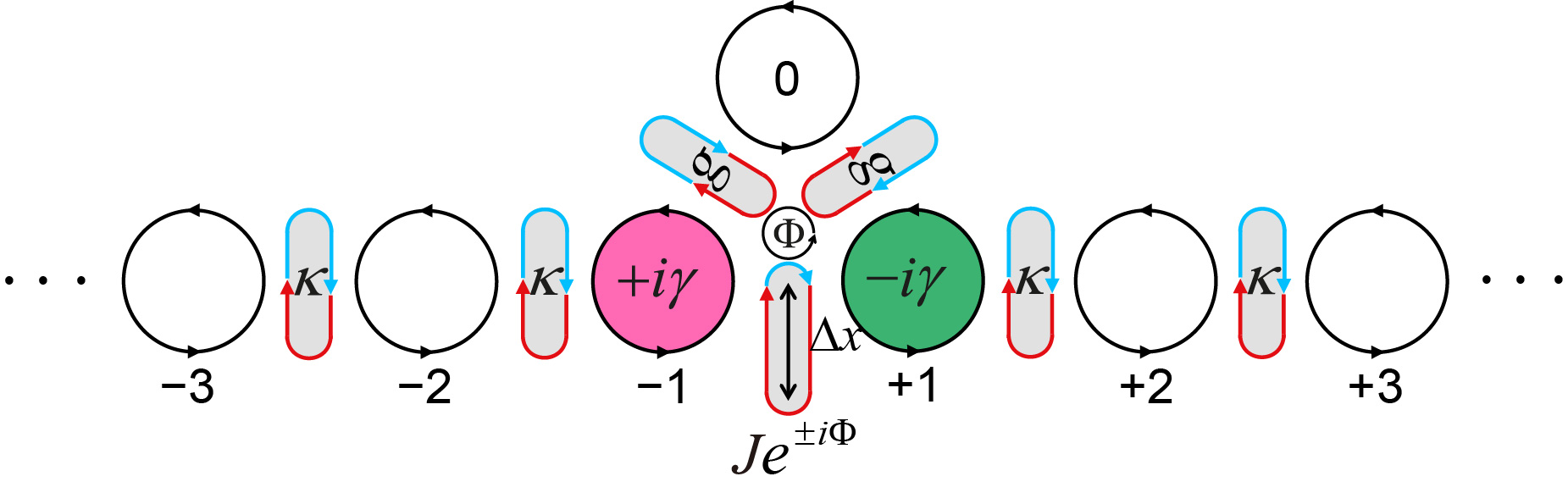}
\caption{Schematic of the coupled resonators AB interferometer. Uniformly
coupled resonator array with the embedded three-coupled-resonator. The gain
(pink) and loss (green) in the resonators are balanced. The passive
resonators are in white. The primary resonators (ring shape) are coupled
through the auxiliary resonators (stadium shape). The photons circling in the primary and auxiliary resonators are in opposite directions.}
\label{fig1}
\end{figure}

The system is a two-arm AB interferometer. One arm comprises the indirect
path from resonator $-1$ to resonator $+1$ through resonator $0$, the coupling
strengths are $-g$. The other arm of the interferometer comprises the
direct path from resonator $-1$ to resonator $+1$; the coupling between
resonators $-1$ and $+1$ has a directional hopping phase, represented by $%
-Je^{\pm i\Phi }$~\cite{HafeziIJMP}. This asymmetric coupling
is introduced using the optical path imbalance method illustrated in Fig.~%
\ref{fig1}. Notably, $\Phi =2\pi \Delta x/\lambda $ depends on the path
length difference $2\Delta x$ experienced by photons in the auxiliary
resonator as they travel between resonators $-1$ and $+1$ in opposite
directions (Fig.~\ref{fig1}), where $\lambda $ is the optical wave length.
The resonator supports clockwise and counterclockwise modes; the black
arrows in Fig.~\ref{fig1} represent the counterclockwise mode. The cyan
(red) arrows indicate the optical path lengths of photons tunneling from
left to right (right to left). Photons tunneling between resonators $-1$ and
$+1$ feel a path difference $2\Delta x$ as illustrated in Fig.~\ref{fig1};
this path difference results in an additional phase factor $e^{\pm i\Phi }$
in the coupling process. The additional directional phase factor corresponds
to synthetic magnetic flux $\Phi $ that is enclosed by the
three coupled resonators~\cite{Hafezi}. The synthetic magnetic
flux is gauge invariant. Although the transmission and reflection
coefficients change, their moduli are invariant under gauge transformation.

In the clockwise mode, photons circle in the opposite direction and all
arrows in Fig.~\ref{fig1} are invert. The path differences in the tunneling
process are opposite for the counterclockwise and clockwise modes.
Consequently, opposite synthetic magnetic fluxes $\Phi $ and $%
-\Phi $ are induced. The synthetic magnetic flux $\Phi $ and
its opposite $-\Phi $ lead to identical transmission and reflection
probabilities. Therefore, the lasing at spectral singularities is identical
for the counterclockwise and clockwise modes.

The modal amplitudes of the resonators are described by coupled-mode theory~%
\cite{CMT}, whereas the equation of motion for the leads is
\begin{equation}
i\dot{a}_{j}=\omega _{\mathrm{c}}a_{j}-\kappa a_{j-1}-\kappa a_{j+1},
\end{equation}%
for the left lead $j<-1$ and the right lead $j>1$. The equations of motion for the scattering center are
\begin{eqnarray}
i\dot{a}_{-1} &=&\left( \omega _{\mathrm{c}}+i\gamma \right)
a_{-1}-ga_{0}-Je^{i\Phi }a_{1}-\kappa a_{-2},  \label{c1} \\
i\dot{a}_{0} &=&\omega _{\mathrm{c}}a_{0}-ga_{-1}-ga_{1},  \label{c2} \\
i\dot{a}_{1} &=&\left( \omega _{\mathrm{c}}-i\gamma \right)
a_{1}-ga_{0}-Je^{-i\Phi }a_{-1}-\kappa a_{2}.  \label{c3}
\end{eqnarray}%
The equations of motion are Schr\"{o}dinger-like equations. The dynamics of
photons in the synthetic magnetic flux described by these equations of
motion are equivalent to those of electrons in magnetic flux in condensed
matter physics.

Notably, the interferometer is $\mathcal{PT}$-symmetric in the presence of synthetic magnetic flux. The parity operator $\mathcal{P}$
satisfies $\mathcal{P}\hat{a}_{-j}^{\dagger }\mathcal{P}^{-1}=\hat{a}%
_{j}^{\dagger }$ and $\mathcal{P}\hat{a}_{j}\mathcal{P}^{-1}=\hat{a}_{-j}$.
The time-reversal operator $\mathcal{T}$ satisfies $\mathcal{T}i\mathcal{T}%
^{-1}=-i$. $\hat{a}_{j}^{\dagger }$ ($\hat{a}_{j}$) is the creation
(annihilation) operator for the resonator $j$.

\section{Spectral singularities and lasing}

\label{AsymmetricLasing} {%
In a finite-size $\mathcal{PT}$-symmetric
non-Hermitian system, the system might experience a phase transition from an
exact $\mathcal{PT}$-symmetric phase to a broken-$\mathcal{PT}$-symmetric
phase at the exceptional points. The exceptional points are points that the finite-size
non-Hermitian system is defective, where coalesced eigenstates appear. Another
type of singularities are spectral singularities in infinite-size
non-Hermitian systems. Spectral singularities are singular points in a
non-Hermitian scattering system at which eigenstates for a continuous
spectrum is incomplete~\cite{AliPRL}. The biorthonornal basis vanishes at the
singularities of non-Hermitian systems~\cite{AliSSJPA09}. In non-Hermitian optical
scattering systems, coherent perfect absorption or lasing occurs at the
spectral singularities of $\mathcal{PT}$-symmetric systems$~$\cite%
{CPA,Science2011,CPAexp,SL82,YDChong106}. The stationary scattering wave
function is that of an incoming wave for a perfect absorber, where outgoing
waves vanish [Fig.~\ref{fig2}(a)]; By contrast, lasing corresponds to a
stationary scattering wave function that constitutes outgoing waves only
[Fig.~\ref{fig2}(b-d)]. Unidirectional perfect absorption has been proposed
at spectral singularities in dissipative systems that are non-$\mathcal{PT}$%
-symmetric~\cite{LXQ,RamezaniUPA,JLSR2}. The $\mathcal{PT}$-symmetric
scattering system considered in this paper is a non-Hermitian two-arm AB
interferometer. We are interested in how spectral singularities and lasing
are affected by the synthetic magnetic flux that causes interference between
the two arms.

\begin{figure}[tbp]
\centering\includegraphics[bb=0 0 420 230, width=7.8 cm, clip]{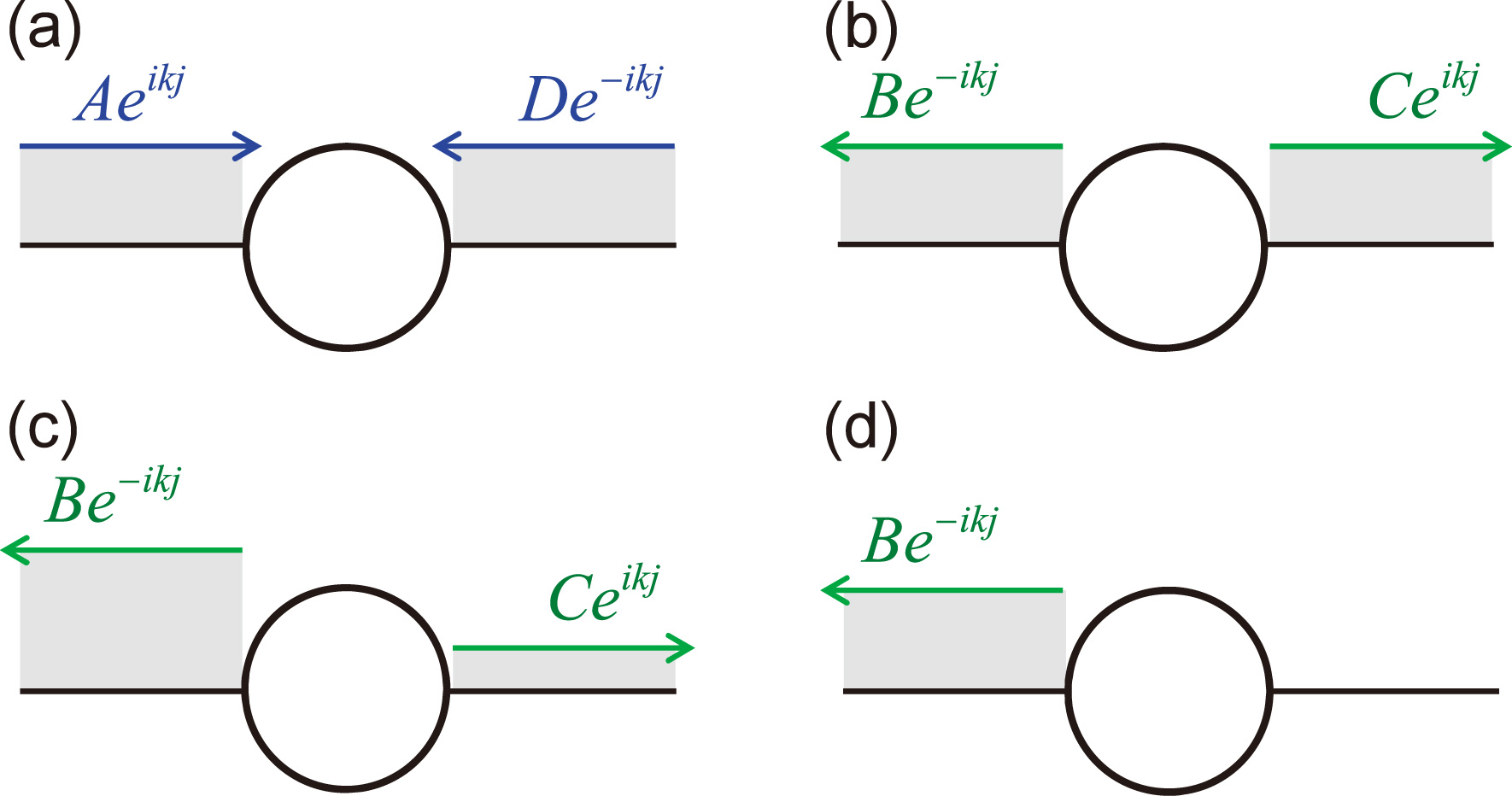}
\caption{The coherent perfect absorbing and lasing. The circles indicate the
non-Hermitian structures. The incoming (outgoing) waves are indicated by the
blue (green) arrows. (a) Coherent perfect absorbing, (b) symmetric lasing,
(c) asymmetric lasing, (d) unidirectional lasing.} \label{fig2}
\end{figure}

The dispersion relation for the coupled
resonators is $\varepsilon =\omega _{\mathrm{c}}-2\kappa \cos k$, where $k$
is the wave vector, $-\pi \leqslant k\leqslant \pi $. The modal amplitude satisfies $a_{j}=f_{j}e^{-i\varepsilon t}$.
From the equations of motion Eqs.~(\ref{c1}-\ref{c3}), we have
\begin{eqnarray}
-gf_{0}-Je^{i\Phi }f_{1}-\kappa f_{-2} &=&\left( E-i\gamma \right) f_{-1},
\label{F1} \\
-gf_{-1}-gf_{1} &=&Ef_{0},  \label{F2} \\
-gf_{0}-Je^{-i\Phi }f_{-1}-\kappa f_{2} &=&\left( E+i\gamma \right) f_{1}.
\label{F3}
\end{eqnarray}%
and $-f_{j-1}-f_{j+1}=Ef_{j}$ for $|j|>1$, where $E=-2\kappa \cos k$. In the elastic scattering
process, the wave functions $f_{j}$ at steady-state are in the form of: $%
f_{j}=Ae^{ikj}+Be^{-ikj}$ ($j<-1$) and $f_{j}=Ce^{ikj}+De^{-ikj}$\ ($j>1$),
where $A$ and $D$ are the amplitudes of the incoming wave, and $B$ and $C$
are the amplitudes of the outgoing wave.

At spectral singularities, the product of the left and right transmissions
is equal to that of the reflections; in other words, $T_{\mathrm{L}}T_{%
\mathrm{R}}=R_{\mathrm{L}}R_{\mathrm{R}}$ holds, where $T$ ($R$) represents
the transmission (reflection) probability. The subscripts indicate either
the left or right input. The scattering coefficients diverge at spectral
singularities for lasing and the wave function consists of outgoing waves
only
\begin{equation}
f_{j}=\left\{
\begin{array}{c}
Be^{-ikj},(j<0) \\
Ce^{+ikj},(j>0)%
\end{array}%
\right. ,
\end{equation}%
where incoming waves $Ae^{ikj}$ $(j<0)$ and $De^{-ikj}$ $(j>0)$ vanish. For
a $\mathcal{PT}$-symmetric interferometer, the transmission or reflection
probability is symmetric. Therefore, symmetric lasing of a $%
\mathcal{PT}$-symmetric interferometer requires not only symmetric reflection and transmission probabilities but also
that these probabilities possess identical values, or in other words, $T_{%
\mathrm{L}}=T_{\mathrm{R}}=R_{\mathrm{L}}=R_{\mathrm{R}}$ [Fig.~\ref{fig2}%
(b)]; otherwise, the lasing is asymmetric [Fig.~\ref{fig2}(c)]. $T_{\mathrm{L%
}}T_{\mathrm{R}}=R_{\mathrm{L}}R_{\mathrm{R}}$\ implies asymmetric lasing
with contrast $\chi=T_{\mathrm{L}}/R_{\mathrm{L}}=R_{\mathrm{R%
}}/T_{\mathrm{R}}$.

The enclosed synthetic magnetic flux affects the interference and spectral singularities. By substituting $f_{-2}=Be^{ik}$, $f_{-1}=B$, $f_{1}=C$, and $f_{2}=Ce^{ik}$
into the equations of motion [Eqs.~(\ref{F1}-\ref{F3})], and then eliminating
$f_{0}$, we obtain\begin{eqnarray}
\left( \frac{g^{2}}{E}-Je^{i\Phi }\right) C &=&\left( E+\kappa e^{ik}-\frac{g^{2}}{E}-i\gamma \right) B,  \label{R1} \\
\left( \frac{g^{2}}{E}-Je^{-i\Phi }\right) B &=&\left( E+\kappa e^{ik}-\frac{g^{2}}{E}+i\gamma \right) C.  \label{R2}
\end{eqnarray}where $E=-2\kappa \cos k$. When $B$ and $C$ are nonvanishing, eliminating $B$ and $C$ from Eqs.~(\ref{R1},~\ref{R2}), we get

\begin{equation}
\left( \kappa e^{-ik}+\frac{g^{2}}{E}\right) ^{2}+\gamma ^{2}=\left( \frac{%
g^{2}}{E}-Je^{i\Phi }\right) \left( \frac{g^{2}}{E}-Je^{-i\Phi }\right) ,
\label{Relation}
\end{equation}%
After simplifying this relation [Eq.~(\ref{Relation})], we obtain the lasing
conditions at spectral singularities as follows:
\begin{eqnarray}
g^{2} &=&2\kappa ^{2}\cos ^{2}k,  \label{g} \\
\cos \Phi &=&\frac{\gamma ^{2}-J^{2}-\kappa ^{2}}{g^{2}}\frac{\kappa }{J}%
\cos k.  \label{Flux}
\end{eqnarray}%
At spectral singularities, the wave function consists of outgoing waves. The
modal amplitudes satisfy $\left\vert B\right\vert ^{2}/\left\vert
C\right\vert ^{2}=\left( \gamma +\kappa \sin k\right) /\left( \gamma -\kappa
\sin k\right) $, which indicates that the lasing is asymmetric in the setup
of the $\mathcal{PT}$-symmetric interferometer.

An extreme case of asymmetric lasing occurs at unidirectional spectral
singularities~\cite{USS}. The lasing occurs in one direction only, where $B$%
\ or $C$ vanishes in the scattering wave function $f_{j}$ [Fig.~\ref{fig2}%
(d)]. In this case, Eqs.~(\ref{R1},~\ref{R2}) reduce to
\begin{eqnarray}
\frac{g^{2}}{2\kappa \cos k}+J\cos \Phi +i\sigma J\sin \Phi  &=&0,
\label{USS1} \\
\kappa \cos k-\frac{g^{2}}{2\kappa \cos k}-i\left( \kappa \sin k+\sigma
\gamma \right)  &=&0.  \label{USS2}
\end{eqnarray}with $\sigma =+1$ ($-1$)$\ $for $B=0$ ($C=0$). Equations~(\ref{USS1},~\ref{USS2}) indicate that the unidirectional spectral singularities place
additional constrains on the parameters of the coupled resonator system. The real and imaginary parts of the left side of Eqs.~(\ref{USS1},~\ref{USS2}) vanish, respectively.

Equation~(\ref{USS1}) yields that the synthetic magnetic flux must be:
\begin{equation}
\Phi =n\pi ,\text{ }(n\in
\mathbb{Z}).  \label{a}
\end{equation}Equation~(\ref{USS2}) yields:
\begin{equation}
\gamma ^{2}=\kappa ^{2}\sin ^{2}k,g^{2}/2=\kappa ^{2}\cos ^{2}k.  \label{c}
\end{equation}
Associated Eq.~(\ref{USS1}) and Eq.~(\ref{USS2}), we notice that the coupling, the synthetic magnetic flux, and the gain or loss rate are related to the wave vector as follows:
\begin{equation}
J\cos \Phi =-\kappa \cos k.  \label{b}
\end{equation}

Notably, $\cos ^{2}\Phi =1$ for $\Phi =n\pi $, $(n\in
%TCIMACRO{\U{2124} }%
%BeginExpansion
\mathbb{Z}
%EndExpansion
)$. Thus, the coupling strengths satisfy $g^{2}=2J^{2}$ and the parameters
take the forms of a circle and an ellipse in parameter space,
\begin{equation}
J^{2}+\gamma ^{2}=\kappa ^{2},
\end{equation}%
and%
\begin{equation}
g^{2}/2+\gamma ^{2}=\kappa ^{2}.
\end{equation}

\begin{figure}[tb]
\centering
\includegraphics[bb=0 0 370 512, width=8.8 cm, clip]{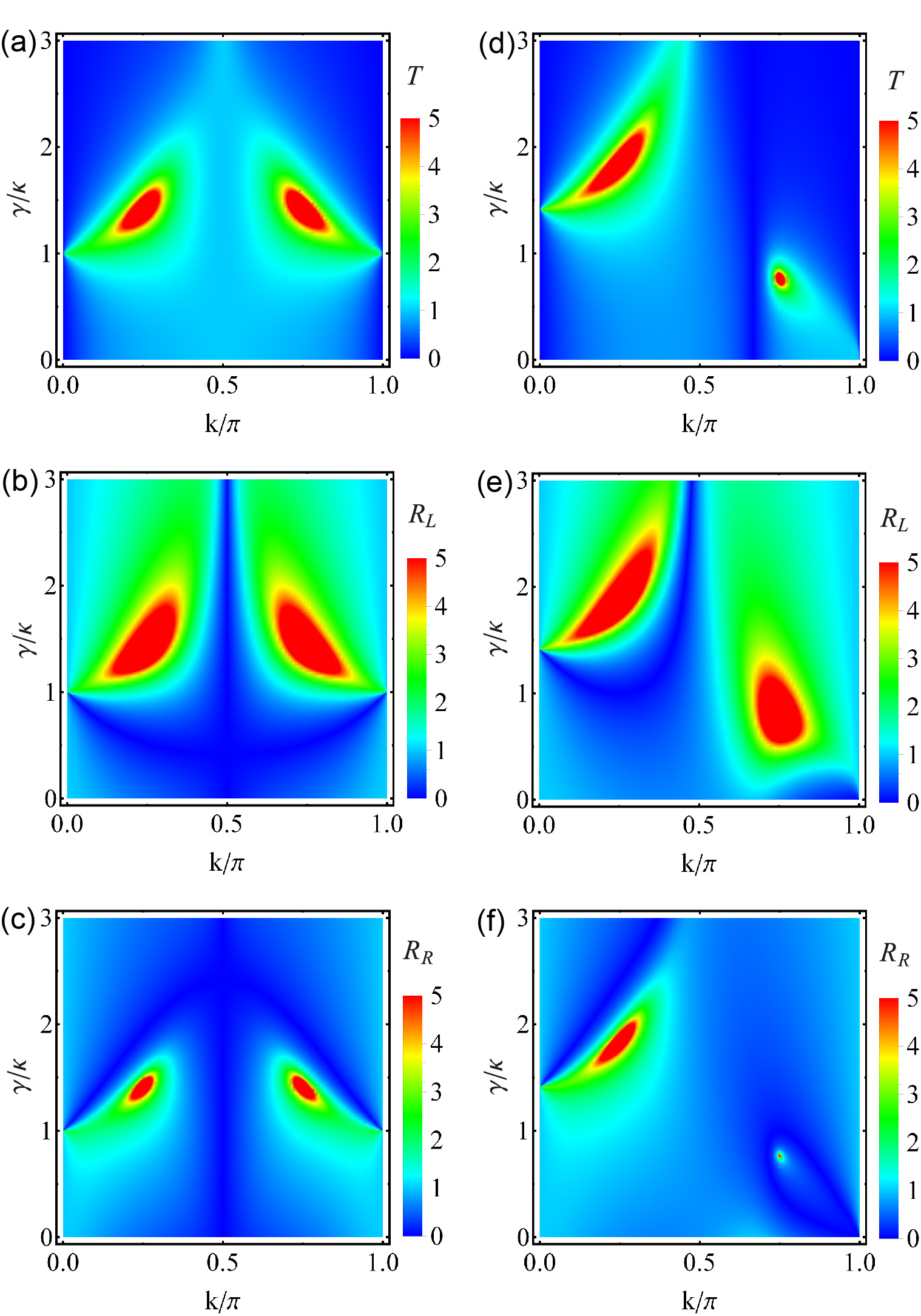}
\caption{Transmission and reflection probabilities. (a-c) $\Phi =\protect\pi /2$ and (d-f) $\Phi =0$. (a, d) Symmetric transmissions $T$, (b, e) left
reflection $R_{\mathrm{L}}$, and (c, f) right reflection $R_{\mathrm{R}}$.} %
\label{fig3}
\end{figure}

At unidirectional spectral singularities, the symmetric transmission is finite; the reflection is infinite for one input direction
and zero for the other input direction. The stationary wave function for $%
\gamma =\kappa \sin k$ is%
\begin{equation}
f_{j}=\left\{
\begin{array}{c}
e^{-ikj},(j<0) \\
0,(j>0)%
\end{array}%
\right. ,
\end{equation}%
and the stationary wave function\ for $\gamma =-\kappa \sin k$ is
\begin{equation}
f_{j}=\left\{
\begin{array}{c}
0,(j<0) \\
e^{+ikj},(j>0)%
\end{array}%
\right. .
\end{equation}%
The stationary wave function for unidirectional lasing implies that the
lasing is directional, and this directionality is generated from the gain
resonator as a reflection toward the lead to which it is connected. By
contrast, the reflection in the other direction is perfectly absorbed by the
loss resonator. The clockwise and counter clockwise modes experience opposite synthetic magnetic fluxes. From the conditions of spectral singularities, we notice that they both lase at spectral singularities for fixed device parameters.

\section{Uniformly coupled $\mathcal{PT}$-symmetric AB interferometer}

\label{EG} As an illustrative example, we investigate the AB interferometer
in the case of $J=g=\kappa $. For the left input with wave vector $k$, the
scattering wave function in the left lead is $f_{j}=e^{ikj}+r_{\mathrm{L}%
}e^{-ikj}$ ($j<-1$) and the transmission wave in the right lead is $f_{j}=t_{%
\mathrm{L}}e^{ikj}$ ($j>1$). By contrast, for the right input, the input
wave function in the right lead is $f_{j}=e^{-ikj}+r_{\mathrm{R}}e^{ikj}$ ($%
j>1$) and the transmission wave in the left lead is $f_{j}=t_{\mathrm{R}%
}e^{-ikj}$\ ($j<-1$). $t_{\mathrm{L}(\mathrm{R})}$ ($r_{\mathrm{L}(\mathrm{R}%
)}$) represents the transmission (reflection) coefficient for the left
(right) input. The transmission and reflection probabilities are $T_{\mathrm{%
L}(\mathrm{R})}=|t_{\mathrm{L}(\mathrm{R})}|^{2}$ and $R_{\mathrm{L}(\mathrm{%
R})}=|r_{\mathrm{L}(\mathrm{R})}|^{2}$. Substituting the wave function $%
f_{j} $ into the equations of motion, the scattering coefficients $r_{%
\mathrm{L},\mathrm{R}}$ and $t_{\mathrm{L},\mathrm{R}}$ are obtained as
follows,
\begin{eqnarray}
t_{\mathrm{L}} &=&\frac{i\sin k\left( 1+e^{-i\Phi }2\cos k\right) }{%
ie^{-2ik}\sin k+\left( 1-\gamma ^{2}\right) \cos k+\cos \Phi }, \\
r_{\mathrm{L}} &=&\frac{\gamma \sin \left( 2k\right) -\left( 1-\gamma
^{2}\right) \cos k-\cos \Phi }{ie^{-2ik}\sin k+\left( 1-\gamma ^{2}\right)
\cos k+\cos \Phi },
\end{eqnarray}%
\begin{eqnarray}
t_{\mathrm{R}} &=&\frac{i\sin k\left( 1+e^{i\Phi }2\cos k\right) }{%
ie^{-2ik}\sin k+\left( 1-\gamma ^{2}\right) \cos k+\cos \Phi }, \\
r_{\mathrm{R}} &=&\frac{-\gamma \sin \left( 2k\right) -\left( 1-\gamma
^{2}\right) \cos k-\cos \Phi }{ie^{-2ik}\sin k+\left( 1-\gamma ^{2}\right)
\cos k+\cos \Phi }.
\end{eqnarray}

From these expressions of the scattering coefficients, we find that the
transmission is symmetric, or in other words, $T=T_{\mathrm{L}%
}=T_{\mathrm{R}}$. Figure~\ref{fig3} illustrates the transmission and
reflection probabilities. In the region $k\in (0,\pi )$, the scattering
probability is symmetric at $k=\pi /2$ for a synthetic magnetic flux of $n\pi +\pi /2$ ($n\in
%TCIMACRO{\U{2124} }%
%BeginExpansion
\mathbb{Z}
%EndExpansion
$) [Fig.~\ref{fig3}(a-c)], where the scattering spectra are
symmetric about the resonance frequency; otherwise (for example, at $\Phi =0$%
), the scattering spectra are asymmetric about $k=\pi /2$ [Fig.~\ref{fig3}(d-f)].
For $\Phi =\pi $, the transmission and reflection probabilities are mirror
reflection symmetric about $k=\pi /2$ at $\Phi =0$. When $k=\pi /2$, the
reflection and transmission are symmetric and $\gamma $%
-independent, $T_{\mathrm{L}}=T_{\mathrm{R}}=(\cos ^{2}\Phi +1)^{-1}$ and $%
R_{\mathrm{L}}=R_{\mathrm{R}}=(\cos ^{2}\Phi +1)^{-1}\cos ^{2}\Phi $. The synthetic magnetic flux alters the reflection zeros whereas
transmission zeros occur only when $\Phi =n\pi $ ($n\in
%TCIMACRO{\U{2124} }%
%BeginExpansion
\mathbb{Z}
%EndExpansion
$). For an input wave vector $k\neq \pi /2$, the symmetric reflection is broken in the presence of nonzero $\gamma $\ even at spectral
singularities because the two arms of the interferometer are imbalanced. The
reflectionless transmission is unidirectional and synthetic magnetic flux dependent.

\begin{figure}[tb]
\centering
\includegraphics[bb=0 0 545 250, width=8.8 cm, clip]{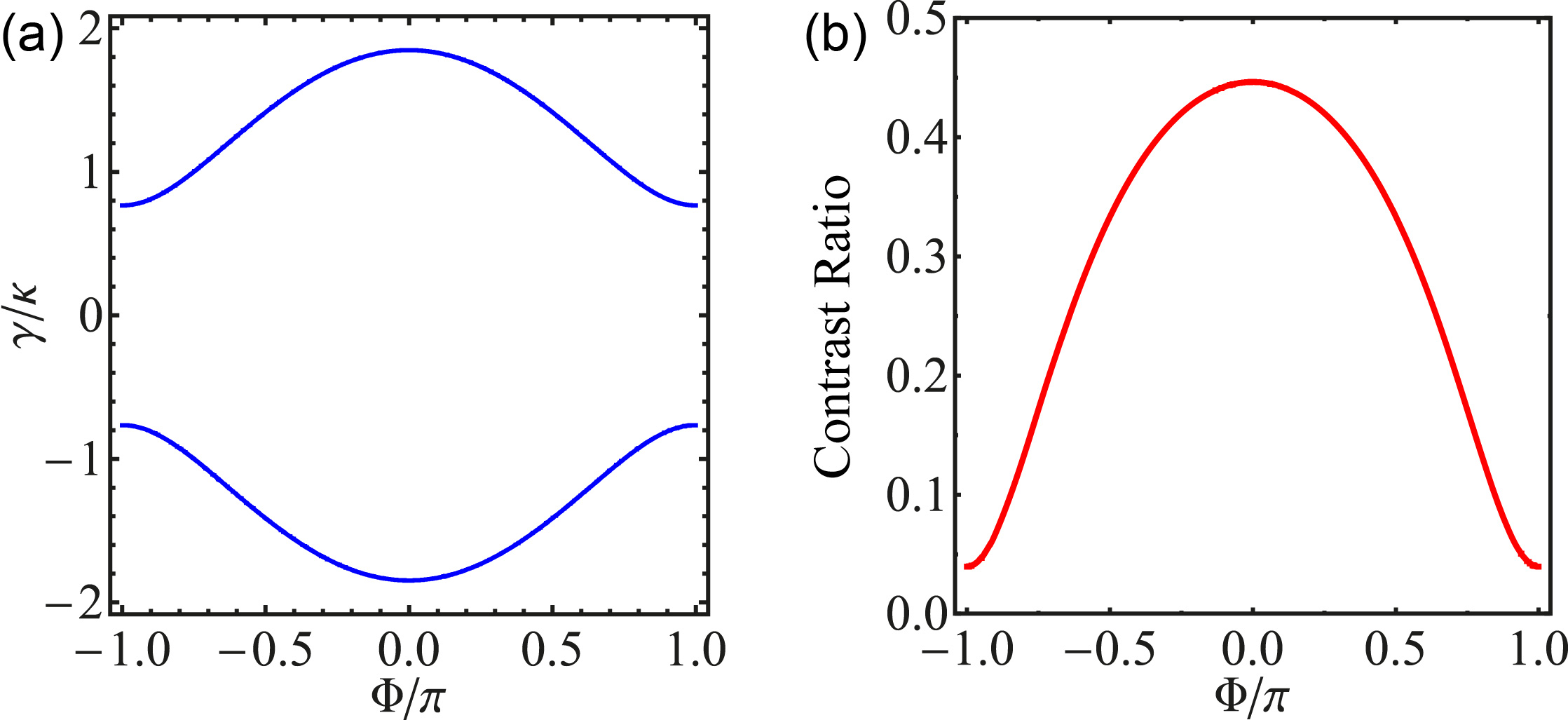}
\caption{(a) Spectral singularities in the $\Phi $ and $\protect\gamma $
parameter space for $k=\protect\pi /4$. The
curves indicate the place where scattering coefficients diverge. Spectral
singularities appear in the region $\protect\sqrt{2-\protect\sqrt{2}}\leqslant
|\protect\gamma /\protect\kappa |\leqslant \protect\sqrt{2+\protect\sqrt{2}}$. (b) The lasing contrast ratio at spectral singularities.}
\label{fig4}
\end{figure}

\begin{figure*}[thb]
\centering
\includegraphics[bb=0 0 550 150, width=17.8 cm, clip]{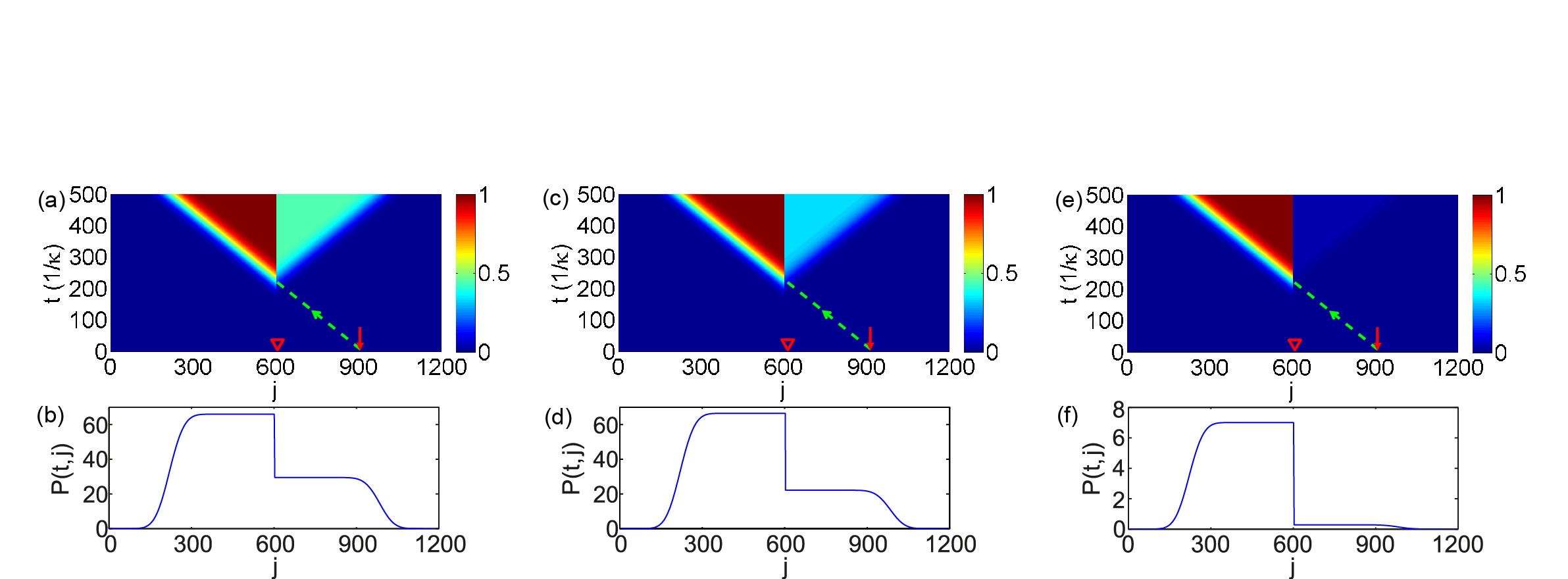}
\caption{Numerical simulations of lasing wave intensity for
(a, b) $\Phi =0$, (c, d) $\Phi =\protect\pi /2$, and (e, f) $\Phi =\protect%
\pi $. (a, c, e)
Density plots of the intensity $P(t,j)/h$, where $h$ is the left-traveling
wave height as shown in (b, d, f) on their left halves. (b, d, f)
Configuration of the lasing wave intensity $P(t,j)$ at $t=500/\protect\kappa
$. The three coupled resonators embedded in the center of the resonator
array are indicated by the red triangles. A Gaussian wave packet with $%
\protect\alpha =0.02$ and a wave vector $k_{\mathrm{c}}=\protect\pi /4$
centered at $N_{\mathrm{c}}=900$ (indicated by the red arrow) is initially
travelling left. The trajectory of the wave packet center is indicated by
the dashed green line. An asymmetric wave is emitted after reaching the
center at approximately $t=210/\protect\kappa $.}
\label{fig5}
\end{figure*}
Divergence of the scattering coefficients occurs only for wave vectors $%
k=\pi /4$ and $3\pi /4$ [obtained from Eq.~(\ref{g})]. Correspondingly,
scattering coefficient divergence emerges when the gain and loss rates
satisfy $\gamma ^{2}=2\pm \sqrt{2}\cos \Phi $ [obtained from Eq.~(\ref{Flux}%
)]. At $k=\pi /4$, the transmissions ($t_{\mathrm{L}}$ and $t_{\mathrm{R}}$)
are larger than zero under any balanced gain and loss $\gamma $ or when the synthetic magnetic flux $\Phi $ is enclosed. Divergence of the
scattering coefficients occurs at
\begin{equation}
\gamma ^{2}=\sqrt{2}\cos \Phi +2,
\end{equation}%
and spectral singularities appear in the region $\sqrt{2-\sqrt{2}}\leqslant
|\gamma /\kappa |\leqslant \sqrt{2+\sqrt{2}}$; the spectral singularity
curves are shown in Fig.~\ref{fig4}(a). At spectral singularities, the
scattering center acts as a wave emitter~\cite{PWang} and the lasing is
asymmetric. The ratio of left-going lasing to right-going lasing, which is
obtained by comparing the transmission and reflection coefficients, is
plotted in Fig.~\ref{fig4}(b). A ratio lower than one indicates that the
left side, which includes the gain resonator, has a higher lasing rate.
Notably, the three embedded coupled resonators act as a wave emitter at
spectral singularities. The lasing is asymmetric, and the contrast ratio of
right-travelling wave emission to left-travelling wave emission is%
\begin{equation}
\chi\left( \Phi \right) =\frac{\sqrt{2\sqrt{2}\cos \Phi +4}-1%
}{\sqrt{2\sqrt{2}\cos \Phi +4}+1},
\end{equation}%
which is obtained by comparing the transmission and reflection
probabilities. The gain resonator is on the left in this system, and thus
left-travelling wave emission is stronger than right-travelling wave
emission and the contrast is less than 1. For example, at $\gamma =\sqrt{2}$
and $\Phi =\pi /2$, we have $T_{\mathrm{L}}/R_{\mathrm{L}}=1/3$ and $T_{%
\mathrm{R}}/R_{\mathrm{R}}=3$. Notably, $T_{\mathrm{L}}T_{\mathrm{R}}=R_{%
\mathrm{L}}R_{\mathrm{R}}$ holds independent of the synthetic magnetic flux at spectral singularities, indicating that the asymmetry of
the lasing is input direction independent, whereas the lasing intensity is
input direction dependent. The contrast $T_{\mathrm{L}}/R_{\mathrm{L}}$
decreases as the synthetic magnetic flux $\Phi $ increases
from $0$ to $\pi $; therefore, the asymmetry of the lasing increases.

A signature of spectral singularities in a wave packet scattering process is
wave emission. We numerically simulate a right input wave emission process
in the AB interferometer. The initial excitation is a normalized Gaussian wave packet of $|\Psi \left( 0, j\right)
\rangle =(\sqrt{\pi }/\alpha )^{-1/2}\sum_{j}e^{-(\alpha ^{2}/2)\left( j-N_{%
\mathrm{c}}\right) ^{2}}e^{ik_{\mathrm{c}}j}\left\vert j\right\rangle $
centered at site $N_{\mathrm{c}}$, where $j$ indicates the label of the
resonator in the array and $k_{\mathrm{c}}$ is the wave vector of the
Gaussian wave packet. At spectral singularities, lasing occurs toward both
sides after the wave packet reaches the embedded resonators, and forms an
asymmetric plateau described by an error function~\cite{PWang}. The lasing wave intensity increases linearly with time, being proportional to the wave propagating velocity $2\kappa\sin{k}$ and the plateau heights. The time evolution of the Gaussian wave packet is $|\Psi \left( t,j\right) \rangle
=e^{-iHt}|\Psi \left( 0,j\right) \rangle $, where $H$ is Hamiltonian of the $1200$-site system including both the trimer scattering center $H_{\mathrm{c}}$ and two finite
leads connected to the center. In Fig.~\ref{fig5}(a,c,e), a Gaussian wave packet is
initially centered at $N_{\mathrm{c}}=900$, the wave vector is $k_{\mathrm{c}%
}=\pi /4$, and the packet moves from right to left at a velocity of $\sqrt{2}%
\kappa $. To emphasis the different asymmetries, the density plots of intensity $P\left(t, j\right)/h$ are depicted for comparison, where $P\left( t, j\right) =\langle \Psi \left(
t, j\right) |\Psi \left( t, j\right) \rangle $ is the intensity of the wave packet and $h$ is the left-travelling wave plateau height in Fig.~\ref{fig5}(b,d,f). At $t\approx 210/\kappa $, the Gaussian wave packet reaches the scattering
center and the intensity begins to increase. Figure~\ref{fig5}%
(b,d,f) depicts the intensity $P\left(t, j\right)$}} of the
lasing at $t=500/\kappa $ and clearly displays the asymmetric plateau. The
left-travelling wave for the right input and the right-travelling wave for
the left input induce the same transmitted plateau height $h$, equal to $3%
\sqrt{\pi }/(4\alpha )$, $3\sqrt{\pi }/(4\alpha )$, and $\sqrt{\pi /10}%
/(4\alpha )$ for $\Phi =0$, $\pi /2$, and $\pi $, respectively [left
plateaus displayed in Fig.~\ref{fig5}(b,d,f)].

Moreover, after the wave packet is scattered by the embedded coupled
resonators, the height of the left-travelling wave is $h\chi^{-1}\left( \Phi \right) $ and that of the right-travelling wave is $h$ for
the left input. By contrast, the height of the left-traveling wave is $h$
and that of the right-travelling wave is $h\chi\left( \Phi
\right) $ for the right input. The contour plots in Fig.~\ref{fig5}(a,c,e)
indicate that the asymmetry is enhanced as the synthetic magnetic flux is increased from $0$ to $\pi $, where the left-travelling
wave emission plateau heights are renormalized to 1.

\section{Conclusion}

\label{Summary} A $\mathcal{PT}$-symmetric two-arm AB interferometer where
three coupled resonators are embedded in a uniformly coupled resonator array
is proposed in this paper. Synthetic magnetic flux is enclosed by the three
coupled resonators using the path length imbalance method. The synthetic magnetic flux acts as a new degree of freedom,
controlling light interference but not breaking the $\mathcal{PT}$ symmetry.
In the two-arm AB interferometer, the $\mathcal{PT}$ symmetry protects the symmetric transmission;
however, the reflection is asymmetric due to the non-Hermitian gain and loss. The conditions for symmetric, asymmetric, and unidirectional lasing at spectral
singularities are discussed. The interplay between the synthetic magnetic flux and non-Hermiticity controls the
lasing at spectral singularities; the lasing is proven asymmetric. The asymmetric lasing at spectral singularities and the
reflectionless transmission both vary with the enclosed synthetic magnetic flux. We demonstrate a uniformly coupled AB interferometer as an illustrative
example and show that the asymmetry of lasing is enhanced as the synthetic magnetic flux is increased. Our results could be
useful in the design of optical control and lasing devices, and may
facilitate the application of $\mathcal{PT}$-symmetric metamaterials.

\section*{Acknowledgements}

We acknowledge the support of National Natural Science Foundation of China
(Grant No. 11605094) and the Tianjin Natural Science Foundation (Grant No.
16JCYBJC40800).

%%%%%%%%%%%%%%%%%%%%%%% References %%%%%%%%%%%%%%%%%%%%%%%%%

\end{document}